\documentclass[runningheads]{llncs}
\usepackage[T1]{fontenc}
\usepackage{graphicx}
\usepackage{dcolumn}% Align table columns on decimal point
\usepackage{bm}% bold math
\usepackage{amsmath,amssymb}
\usepackage{hyperref}
\usepackage{wrapfig}
\usepackage[utf8]{inputenc}
\usepackage[varg]{newtxmath}
\usepackage{color}
\newcommand{\bxi}{{\bm{\xi}}}
\newcommand{\bv}{{\bm{v}}}
\newcommand{\mS}{{\mathcal{S}}}
\newcommand{\Kn}{{\mathrm{Kn}}}
\begin{document}
\title{Heat transfer problem of a dense gas described by the
Enskog equation with a modification of the Enskog factor}
\titlerunning{Heat transfer problem of a dense gas}
\author{Shigeru Takata\inst{1}\orcidID{0000-0001-6787-6777} \and
Soma Sakata\inst{1}\orcidID{0009-0004-1164-7157} \and
Masanari Hattori\inst{1}\orcidID{0000-0002-5482-0210}}
\authorrunning{S. Takata et al.}
\institute{Department of Aeronautics \& Astronautics,
Kyoto University, Kyoto 615-8540, Japan\\
}
\maketitle              % typeset the header of the contribution
\begin{abstract}
Heat transfer in a dense gas between two parallel plates is studied for assessing the quantitative difference of two variants of the Enskog equation, i.e., the original Enskog equation and the Enskog equation with a modified Enskog factor recently proposed in Phys. Rev. E \textbf{111}, 065108 (2025). The main advantage of the latter is that the H theorem, which the former lacks, has been established. The influence of the modification is assessed by comparing the macroscopic quantities in detail. It is found that the impact of this modification is quite limited and the two variants reasonably agree as a whole.
\keywords{Kinetic theory \and Enskog equation \and Enskog factor.}
\end{abstract}

\section{Introduction \label{sec:Intro}}

Recently a study of the dense gas behavior through the kinetic theory attracts attentions, 
coming from the increasing demands for applications to micro-, submicro-, and even smaller systems. 
As the mean free path of molecules and the system size simultaneously decrease, 
the molecular size and the volume fraction of molecules become not necessarily negligible, 
and accordingly the Enskog equation \cite{E72,S16} comes into play in place of the Boltzmann equation.
Although the Enskog equation with its original form (original Enskog equation, 
OEE for short) has been successfully applied to fundamental physical problems, 
thanks to the developments of its numerical methods \cite{F97a,MS97a,WZR15},
it does not satisfy the Boltzmann’s H theorem, unfortunately.
In the meantime, we have recently proposed \cite{TT25} a variant of the Enskog equation 
that slightly modifies the Enskog factor in the collision integral (EESM for short), 
for which the H theorem has been established.
This new variant is currently in the initial stages of quantitative evaluation 
for application to physical problems.

In the present work, we revisit the time evolutional heat transfer problem 
between two parallel plates \cite{F99,HTT22}.
On the basis of both OEE and EESM,
we numerically study the gas behavior
triggered by a sudden temperature change of one plate 
and evaluate the influence of modifying the Enskog factor.

%\newpage

\section{Problem and Formulation \label{sec:Prob}}

\begin{wrapfigure}{r}{0.45\textwidth}
\centering
%\vspace*{-1.5\baselineskip}
\includegraphics[width=0.45\textwidth]{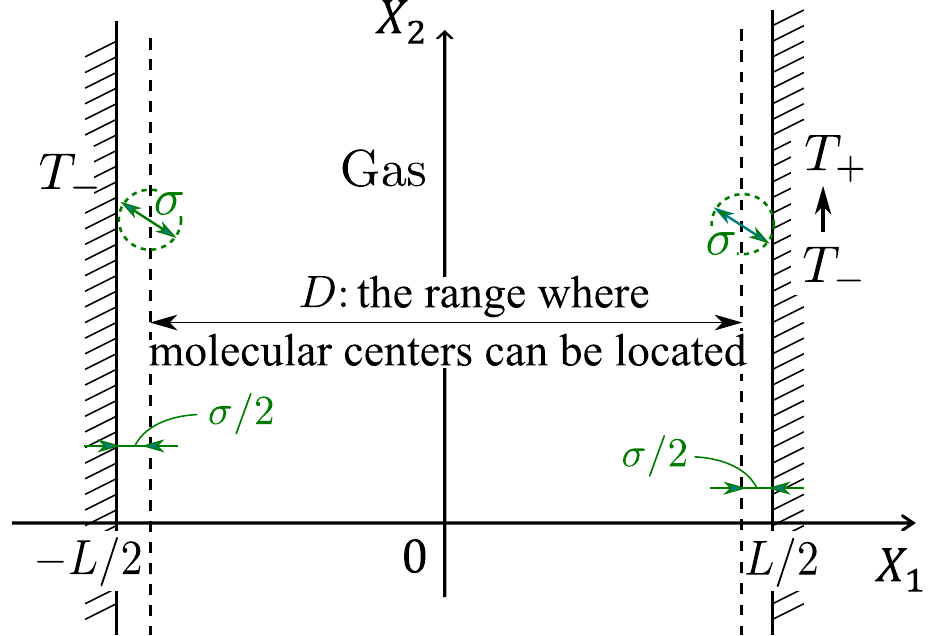}
\caption{Problem setting\label{fig:setting}}
\vspace*{-1.5\baselineskip}
\end{wrapfigure}

Consider a dense gas between two parallel plates at rest located at $X_1=\pm L/2$.
The gas is in thermal equilibrium with the plates kept at a common temperature $T_-$.
At time $t=0$, the temperature of the plate at $X_1=L/2$ is instantaneously changed
from $T_-$ to $T_+$, see Fig.~\ref{fig:setting}.
We will investigate the time-dependent behavior of the gas triggered by this change
under the following assumptions.
(i) the gas molecules are hard spheres of a common diameter $\sigma$ and a mass $m$,
(ii) the behavior of the gas is described by the Enskog equation,
(iii) the gas obeys the Carnahan--Starling equation of state (EoS) \cite{CS69} at a uniform equilibrium state,
(iv) the state of the gas does not change in the $X_2$- and $X_3$-directions,
and (v) the molecules are diffusely reflected on the plates.

Let $D=\{X_1| -(L-\sigma)/2 < X_1 < (L-\sigma)/2\}$, 
where the center of gas molecules can be located.
This $D$ is different from $D^\sharp=\{X_1| -L/2 < X_1 < L/2\}$,
which denotes the spatial domain between the plates.
Let $t$ and $\bm{\xi}$ be a time and a molecular velocity, respectively.
Then, denoting the one-particle velocity distribution function (VDF) of gas molecules
by $f(t,X_1,\bm{\xi}$), 
the Enskog equation for the present spatially one-dimensional problem
is written as
\begin{align}
 \frac{\partial f}{\partial t}+\xi_1\frac{\partial f}{\partial X_1}=
 \frac{\sigma^{2}}{m}
&      \int_{\mathbb{S}^2\times\mathbb{R}^3} 
  [{g(X_1^+,X_1)f_{*}^{\prime}(X_1^+)f^{\prime}(X_1)} \notag\\
& -{g(X_1^-,X_1)f_{*}(X_1^-)f(X_1)}] V_{\alpha}\theta(V_{\alpha})d\Omega(\bm{\alpha})d\bm{\xi}_{*},\label{MEE}
%   \quad \mathrm{for\ } ,
\end{align}
\noindent
where $X_1\in D$, $X_1^\pm=X_1\pm\sigma\alpha_1$,
$\bm{\alpha}=(\alpha_1,\alpha_2,\alpha_3)$ is a unit vector, 
$d\Omega(\bm{\alpha})$ is a solid angle element in the direction of $\bm{\alpha}$,
$\theta$ is the Heaviside function, i.e., $\theta(x)=1$ for $x\ge0$ and $0$ otherwise,
and the following notation convention has been used:
\begin{align}
&
f(\cdot)=f(\cdot,\bm{\xi}),\quad f^{\prime}(\cdot)=f(\cdot,\bm{\xi}^{\prime}), \quad
f_{*}(\cdot)=f(\cdot,\bm{\xi}_{*}),\ f_{*}^{\prime}(\cdot)=f(\cdot,\bm{\xi}_{*}^{\prime}), \label{eq:contf}\\
&
 \bm{\xi}^{\prime}=\bm{\xi}+V_{\alpha}\bm{\alpha},\quad
 \bm{\xi}_{*}^{\prime}=\bm{\xi}_{*}-V_{\alpha}\bm{\alpha},
\quad V_{\alpha}=\bm{V}\cdot\bm{\alpha},\quad\bm{V}=\bm{\xi_{*}}-\bm{\xi}.
\end{align}
\noindent
Here and in what follows, the argument $t$ is often suppressed, unless confusion is anticipated. 
The convention \eqref{eq:contf} will be applied only to the quantities that depend on molecular velocity.
The factor $g$ requires detailed explanation, 
which will be provided later in this section. 

The diffuse reflection %boundary 
condition \cite{S07} in the present problem is written as
\begin{equation}\label{eq:bc}
 f(\pm\frac{L-\sigma}{2})=\frac{1}{2\pi (RT_\pm)^{2}}
\Big(
\int_{\xi_{*1}\gtrless 0}|\xi_{*1}|f_*(\pm\frac{L-\sigma}{2}) d\bxi_* 
\Big)
\exp(-\frac{\bxi^2}{2RT_\pm}),\ 
  \xi_{1}\lessgtr 0,\ t>0,
\end{equation}
\noindent
where $R$ is the specific gas constant.

It should be noted that \eqref{MEE} makes sense only 
when the pair of positions $(X_1,X_1^\pm)$ is in $D$.
Hence, $g$ should be considered as a function such that
\begin{equation}
{g}(X_1,Y_1)=\mathsf{g}(X_1,Y_1)\chi_D(X_1)\chi_D(Y_1),
\end{equation}
\noindent
where $\chi_D(x)$ is the indicator function, i.e., $\chi_D(x)=1$ for $x\in D$ and $0$ otherwise,
$\mathsf{g}$ is positive and symmetric with respect to the exchange of positions: 
$\mathsf{g}(X_1,Y_1)=\mathsf{g}(Y_1,X_1)$.
We call $\mathsf{g}$ the Enskog factor in the sequel.
Although some variants of $\mathsf{g}$ have been proposed in the literature (e.g.,\cite{E72,VE73,BB18,BB19}),
we focus on two variants in the present paper:
the original form by Enskog \cite{E72}
\begin{subequations}
\begin{equation}\label{eq:gOEE}
 \mathsf{g}(X_1,Y_1)
=2\mathcal{S}(\frac{4\pi\sigma^3}{3m}\rho(\frac{X_1+Y_1}{2})),
\end{equation}
\noindent
and the form proposed in \cite{TT25}
\begin{equation}
 \mathsf{g}(X_1,Y_1)=\mathcal{S}(\mathcal{R}(X_1))+\mathcal{S}(\mathcal{R}(Y_1)),
\label{eq:gEESM}
\end{equation}
\noindent
where
\begin{equation}
\mathcal{R}(X_1)=\frac{2\pi}{m}\int_D \int_0^\infty\rho(Z_1)
\theta(\sigma-\sqrt{r^2+(Z_1-X_1)^2})rdr dZ_1,
\label{eq:R_def}
\end{equation}
\noindent
and $\rho$ is the mass density defined as
\begin{equation}
\rho=\langle f \rangle,\label{eq:rho_def}
\quad \langle \bullet\rangle\equiv\int_{\mathbb{R}^3}\bullet\ d\bxi.
\end{equation}
\noindent
What we call OEE is the Enskog equation equipped with \eqref{eq:gOEE},
while what we call EESM is that equipped with \eqref{eq:gEESM}.
Moreover, the form of $\mS$ is related to
the EoS under consideration as
\begin{equation}\label{eq:EoS}
p=\rho R T (1+8\eta\mS(8\eta)),\quad \eta=\frac{\pi\sigma^3}{6m}\rho,
\end{equation}
\noindent
in the uniform equilibrium state,
where $p$ is the thermodynamic static pressure
and $T$ is the temperature defined by 
\begin{equation}
 T=\frac{1}{3R\rho}\langle (\bm{\xi}-\bm{v})^2 f \rangle, \label{eq:temp_def}
\quad \bv=\frac{1}{\rho}\langle \bxi f \rangle.
% \label{eq:flow_def}
\end{equation}
\noindent
In the above, $\eta$ is the volume fraction of molecules \cite{S16}
and $\bv=(v_1,v_2,v_3)$ is the flow velocity.
Since the Carnahan--Starling EoS \cite{CS69} is written as
\begin{equation}
 \frac{p}{\rho RT}
=\frac{1+\eta+\eta^2-\eta^3}{(1-\eta)^3}
=1+\frac{2\eta(2-\eta)}{(1-\eta)^3}
,
\label{eq:C-SEoS}
\end{equation}
\noindent
the corresponding $\mS$ is given as
\begin{equation}
 \mathcal{S}(x)=\frac{16(16-x)}{(8-x)^3}. \label{eq:SCS}
\end{equation}
\end{subequations}

In closing this section,
we give the definition of the stress tensor $p_{ij}$ and the heat flow vector $q_i$
for later convenience.
They are expressed as a sum of the kinetic and the collisional contribution:
\begin{equation}
p_{ij}=\langle c_ic_jf\rangle
      +\frac{\sigma^2}{2m}\int_{\Bbb{S}^2\times\Bbb{R}^3\times\Bbb{R}^3}
       \int_0^\sigma\!\!\! \alpha_i\alpha_j V_\alpha^2 \theta(V_\alpha)
       G(X_1,s,\alpha_1;\bxi_*,\bxi)dsd\Omega(\bm{\alpha})d\bxi_*d\bxi,
\end{equation}
\begin{align}
q_{i}=\frac12\langle c_i\bm{c}^2 f\rangle
     +\frac{\sigma^2}{4m}\int_{\Bbb{S}^2\times\Bbb{R}^3\times\Bbb{R}^3}
      \int_0^\sigma & \alpha_i(\bm{c}+\bm{c}_*)\cdot\bm{\alpha} V_\alpha^2\theta(V_\alpha)\notag\\
& \times      G(X_1,s,\alpha_1;\bxi_*,\bxi)dsd\Omega(\bm{\alpha})d\bxi_*d\bxi,
\end{align}
\noindent
where the outer integration with respect to $\bm{\alpha}$, $\bm{\xi}_*$, and $\bm{\xi}$ is over their whole domain, i.e., $\Bbb{S}^2\times\Bbb{R}^3\times\Bbb{R}^3$, 
$\bm{c}=\bm{\xi}-\bm{v}$, $\bm{c}_*=\bm{\xi}_*-\bm{v}$, and
\begin{equation}
G(X_1,s,\alpha_1;\bxi_*,\bxi)=g(X_1+s\alpha_1,X_1+(s-\sigma)\alpha_1)f_*(X_1+(s-\sigma)\alpha_1)f(X_1+s\alpha_1).
\end{equation}

\section{Numerical analysis\label{sec:Numerics}}

\subsection{Numerical method and related data overview\label{sec:outline}}

Numerical computations have been carried out for the dimensionless version of the problem
\eqref{MEE}--\eqref{eq:bc} supplemented by the initial condition
stated in the problem setting.
Because of the sudden change of the plate temperature,
there is a mismatch of the initial condition and the boundary condition at $X_1=(L-\sigma)/2$
and it propagates into the gas as a discontinuity of the velocity distribution function. 
However, we have not made a special care of it in the present work, 
since the present work primarily aim at observing the quantitative difference
of macroscopic quantities between OEE and EESM.

The numerical method adopted here is basically a combination of the second-order upwind (backward) finite-difference method for a spatial (time) derivative%
\footnote{The first-order scheme is used at the initial time step for time derivative and at the spatial grid point next to the boundary of $D$ for the spatial derivative. The latter applies to the molecular velocity direction leaving the boundary.} 
and the fast Fourier spectral method for the collision integral.
The code is essentially the same as that was used in \cite{TSTH26}
improved from the original code for OEE developed in \cite{HTT22}.
For further details, the reader is referred to these references.
The idea of applying the fast Fourier spectral method to the collision integral
can be found in \cite{FMP06} and the references therein.
Its application to the Enskog equation was probably first done in \cite{WZR15}.
Although numerical convergence has been checked with several patterns of grid system,
the results to be presented below are all carried out with 480 nonuniform grid intervals 
(varying from $1.1\times10^{-5}$ to $3.3\times10^{-3}$ multiplied by $1-\sigma/L$)
in $X_1/L$ with finer grid intervals near the boundaries, 512 uniform grid intervals in $\xi_1/\sqrt{2RT_-}$, and 32 uniform grid intervals both in $\xi_2/\sqrt{2RT_-}$ and $\xi_3/\sqrt{2RT_-}$ directions in the finite cubic domain $[-8,8]^3$. 
The time discretization is uniform with the interval $\Delta t=1.0\times10^{-3}t_0$ with $t_0=L/\sqrt{2RT_-}$.

\begin{figure}[t]
\centering
\includegraphics[height=0.313\textwidth]{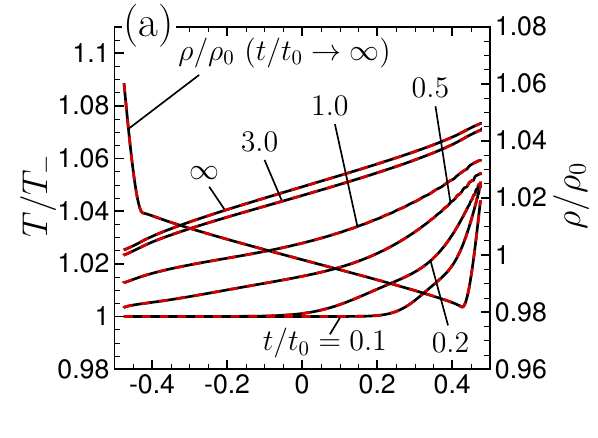} 
\quad
\includegraphics[height=0.313\textwidth]{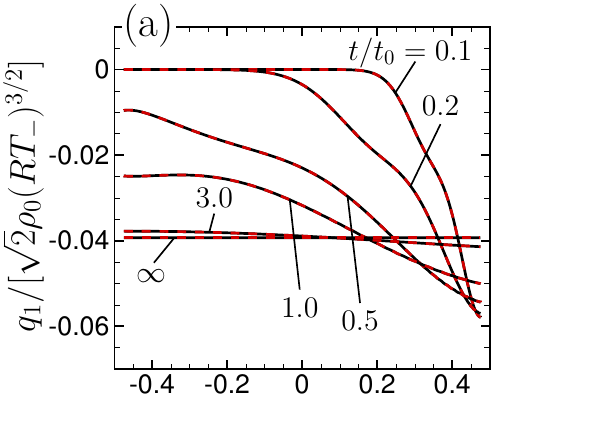} \\
\includegraphics[height=0.313\textwidth]{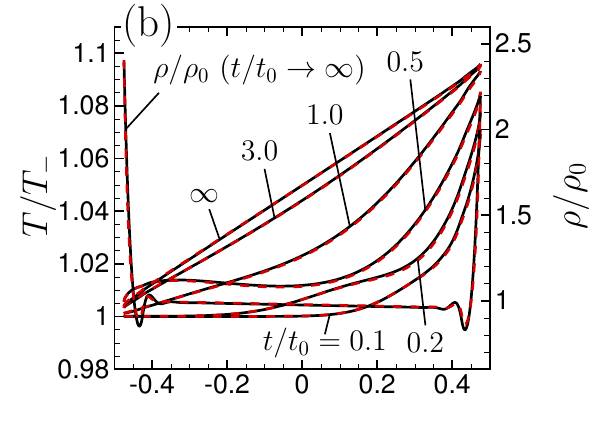} 
\quad
\includegraphics[height=0.313\textwidth]{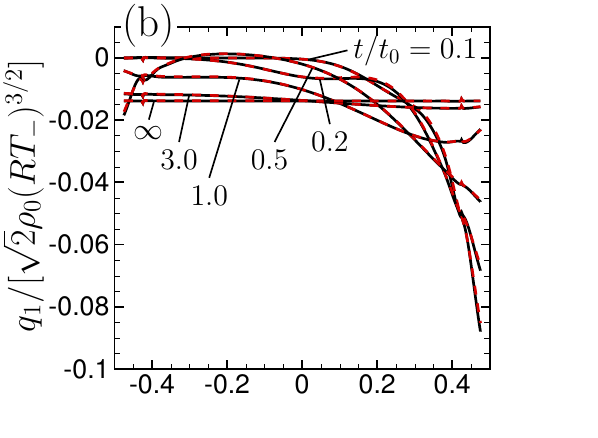} \\
\includegraphics[height=0.313\textwidth]{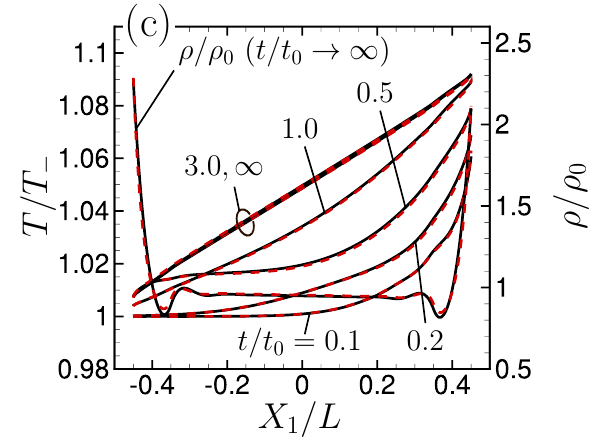} 
\quad
\includegraphics[height=0.313\textwidth]{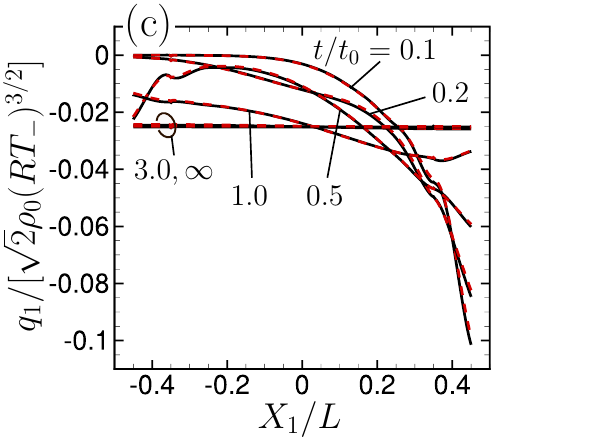} \\
\caption{Time evolution of the temperature $T$ and the density $\rho$ at the final steady state (left column)
and of the heat flow $q_1$ (right column) in the case of $T_+/T_-=1.1$. 
(a) $\sigma/L=0.05$ and $\eta_0=0.01$ ($\Kn=0.5746$). (b) $\sigma/L=0.05$, $\eta_0=0.2$ ($\Kn=0.0168$).
(c) $\sigma/L=0.1$, $\eta_0=0.2$ ($\Kn=0.0335$). 
Solid lines: EESM. Dashed lines: OEE. \label{fig:T}}
\end{figure}

\subsection{Results and discussions\label{sec:results}}

The present problem is characterized by three parameters: 
$T_+/T_-$, $\sigma/L$, and $\eta_0=(\pi\sigma^3/6)(\rho_0/m)$, 
where $\rho_0=\int_D \rho dX_1/(L-\sigma)$.
The Knudsen number $\Kn$ is not an independent parameter,
since it is expressed in terms of $\sigma/L$ and $\eta_0$ as
\begin{equation}\label{eq:Kn}
\Kn\equiv \frac{\ell_0}{L}=\frac{\sigma/L}{12\sqrt{2}\eta_0\mS(8\eta_0)},
\quad 
\ell_0=\frac{1}{2\sqrt{2}\pi\sigma^2(\rho_0/m)\mS(8\eta_0)},
\end{equation}
where $\ell_0$ is the reference mean free path of a molecule.

The time evolution of the temperature $T$ and the density $\rho$ at the final steady state are shown in the left column of Fig.~\ref{fig:T}
for three different pairs of $\eta_0$ and $\sigma/L$ 
in the case of $T_+/T_-=1.1$. 
As is seen, the results based on OEE and EESM
are almost identical over the time evolution process.
The same applies to the time evolution of the heat flow $q_1$,
which is also shown in the right column of Fig.~\ref{fig:T}.
These agreements between OEE and EESM are observed even
for higher volume fraction in our computations.

\begin{figure}[t]
\centering
\includegraphics[width=0.4\textwidth]{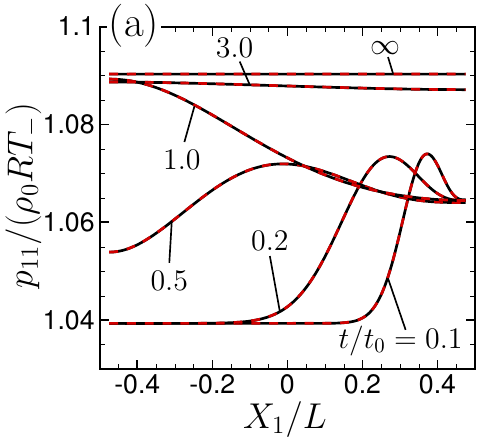} 
\quad
\includegraphics[width=0.4\textwidth]{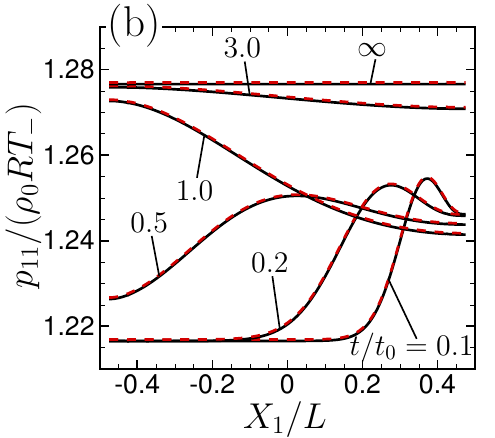} 
\\
\includegraphics[width=0.4\textwidth]{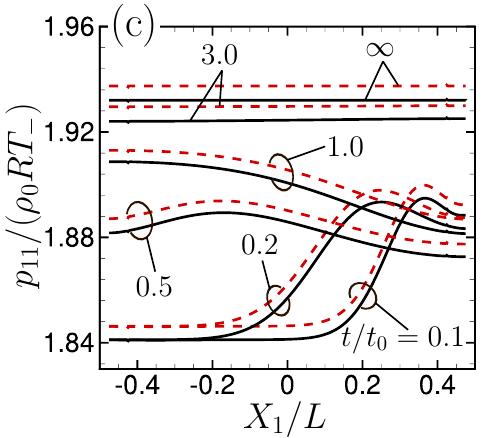} 
\quad
\includegraphics[width=0.4\textwidth]{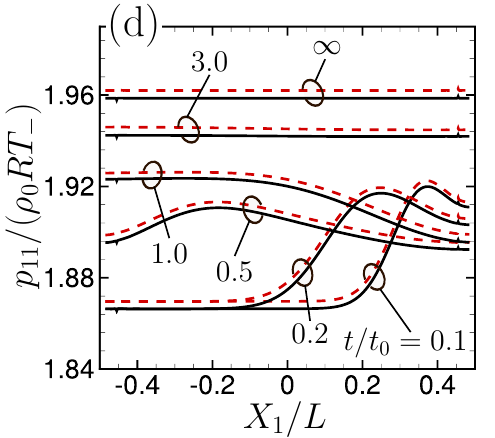} 
%\quad
%\includegraphics[width=0.4\textwidth]{figures/EESMvsOEE_ptot11_t=0.1_0.2_0.5_1.0_3.0_steady_N=120_M1=128_etc_T1=1.1_eta=0.15_sigma=0.01_ver0805_rev2} 
%
\caption{Time evolution of the stress tensor $p_{11}$ in the case of $T_+/T_-=1.1$. 
(a) $\sigma/L=0.05$, $\eta_0=0.01$ ($\Kn=0.5746$). 
(b) $\sigma/L=0.05$, $\eta_0=0.05$ ($\Kn=0.1036$).
(c) $\sigma/L=0.05$, $\eta_0=0.15$ ($\Kn=0.0261$).
(d) $\sigma/L=0.03$, $\eta_0=0.15$ ($\Kn=0.0156$). 
%(d) $\sigma/L=0.01$, $\eta_0=0.15$ ($\Kn=0.0052$). 
Solid lines: EESM. Dashed lines: OEE.\label{fig:P}}
\end{figure}

\begin{figure}[t]
\centering
\includegraphics[width=0.4\textwidth]{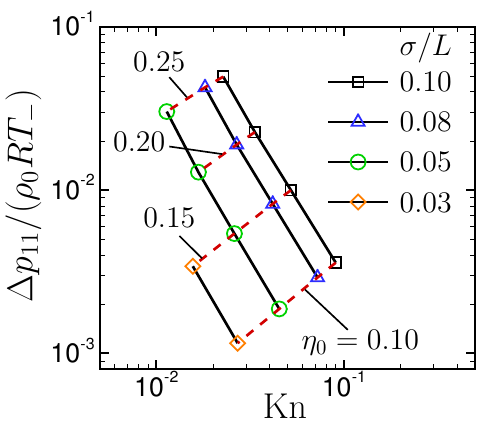}
\caption{The difference of stress tensor $\Delta p_{11}=p_{11}^\mathrm{OEE}-p_{11}^\mathrm{EESM}$ between OEE and EESM at the steady state in the case of $T_+/T_-=1.1$.\label{fig:dP}}
\end{figure}
\begin{figure}[t]\centering
\includegraphics[width=0.42\textwidth]{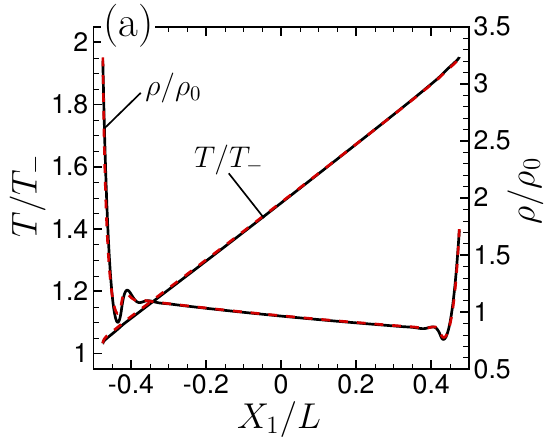} 
\quad
\includegraphics[width=0.42\textwidth]{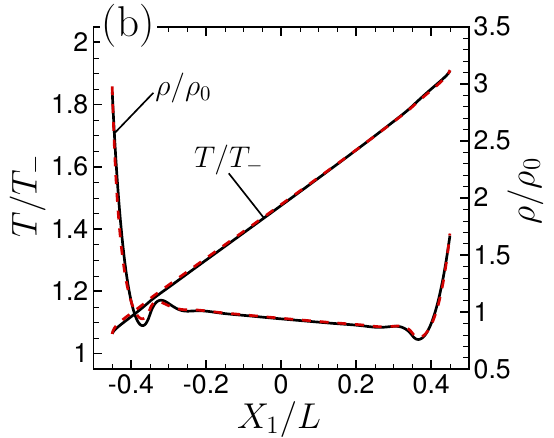} 
\caption{Temperature and density profiles at the final steady state in the case of $T_+/T_-=2$ and
$\eta_0=0.2$. (a) $\sigma/L=0.05$ ($\Kn=0.0168$). (b) $\sigma/L=0.1$ ($\Kn=0.0335$).
Solid lines: EESM. Dashed lines: OEE.\label{fig:T2}}
\end{figure}

However, this is not necessarily the case for the stress tensor $p_{11}$ as shown in Fig.~\ref{fig:P}.
A discernible difference can be observed in the time evolution 
for highly confined cases with large volume fraction,
as shown in Fig.~\ref{fig:P}(c).
In reality, the variation of $p_{11}$ is much smaller than the variations of its kinetic and collisional parts near the boundary. 
The large variations of these two parts mostly cancel each other out, resulting in much smaller variation of $p_{11}$. Accordingly, the discernible difference observed in $p_{11}$ is not truly significant 
relative to the scale of variation of those respective parts. 
The dependence of this difference on $\Kn$ and $\sigma/L$ (or $\eta_0$)
is evaluated based on the constant value achieved at the final steady state,
and the results are shown in Fig.~\ref{fig:dP}.
The difference is proportional to $\Kn^{-2}$ (solid lines in Fig.~\ref{fig:dP}) and 
increases with an increase in $\sigma/L$ and/or $\eta_0$.
It is a little curious that the difference also appears in the regime where both $\sigma/L$ and $\Kn$ are small,
since in this regime EESM and OEE are expected to give the identical results up to the Navier--Stokes level \cite{TT25b} according to the Chapman--Enskog (CE) expansion.
This is likely due to the deviation near the boundary, where the CE expansion does not apply, shifting the entire profile between the plates. The difference is, in general, larger for larger $\sigma/L$ and larger $\eta_0$. 
However, it is also observed that, if $\eta_0$ is fixed, 
the difference in $p_{11}$ decreases as $\Kn$ decreases, see the dashed lines in Fig.~\ref{fig:dP}. 
This implies that the deviation near the boundary also decreases if $\Kn$ and $\sigma/L$ simultaneously decrease with their ratio fixed, namely under the parameter setting consistent with the CE expansion (see \cite{TT25b}).
The decreasing rate is, however, a little worse than $O(\Kn)$, say $O(\Kn^{0.7})-O(\Kn^{0.95})$.

We have also made comparisons with preliminary molecular dynamics results \cite{Kubota}
at the steady states.
As a whole, both the EESM and OEE results agree reasonably with the MD results.
For example, 
the relative error in $p_{11}$ ($q_1$) between EESM and MD is less than 1.3\% (0.7\%)
in the same cases as in Fig.~\ref{fig:T2} that represent rather hard physical settings.
The profile with dimples in density near the boundary is closer between EESM and MD than between OEE and MD. 
The details are, however, omitted here.
 
In closing, the density and temperature at the final steady state for $T_+/T_-=2$ and $\eta_0=0.2$
are shown in Fig.~\ref{fig:T2} as an example of comparisons for extremely severe situations.
It is seen that, in contrast to the rarefied gas, 
the density profile does not change inversely with temperature,
which is considered a characteristic of a non-ideal gas. 
However, it should be also noted that
the density at the boundary is inversely proportional to the temperature at the boundary.
This is a manifestation of the ideal gas behavior there, 
since the collisional contribution to the pressure is suppressed at the boundary.
These features also hold true for the results already shown in Fig.~\ref{fig:T}.

\section{Conclusion}

In the present work, we have numerically studied the behavior of a gas between two parallel plates,
initially in thermal equilibrium with the plates at a common temperature,
induced by a sudden temperature change of one plate. 
The analyses have been carried out on the basis of both the OEE and the EESM, 
with a special interest in the influence of modifying the Enskog factor. 
Overall, the two variants show reasonable agreement, 
except for rather extreme choices of parameters corresponding to a high volume fraction and high confinement. 
The primary difference lies in the depth of the first few density troughs away from the boundary.

\subsubsection{\ackname}
%The present work 
This work is partly supported by Kyoto University Foundation,
by JSPS Grant-in-Aid for Scientific Research(B) (No.~26K00870), and by HPCI system Research Projects (Nos.~hp250004 and hp260026).
A part of computations was conducted %as well 
on SQUID at D3 Center, The University of Osaka.

%\nocite{*}
%\bibliography{aipsamp}% Produces the bibliography via BibTeX.

\end{document}